\documentclass[floatfix,aps,amsmath,nofootinbib,twocolumn,10pt]{revtex4}

\usepackage{graphicx}
\usepackage{bm}

\newcommand{\figurewidth}{3.2in}

\def\({\left(}
\def\){\right)}
\def\[{\left[}
\def\]{\right]}

\def\e{\begin{equation}}
\def\q{\end{equation}}
\def\m{\begin{eqnarray}}
\def\n{\end{eqnarray}}

\begin{document}

\title{How the Dark Energy Can Reconcile \textit{Planck} with Local Determination of the Hubble Constant}

\author{Qing-Guo Huang$^{1,2}$\footnote{huangqg@itp.ac.cn} and Ke Wang$^1$\footnote{wangke@itp.ac.cn}}
\affiliation{$^1$CAS Key Laboratory of Theoretical Physics, Institute of Theoretical Physics, Chinese Academy of Sciences, Beijing 100190, China\\
$^2$School of Physical Sciences, University of Chinese Academy of Sciences, No. 19A Yuquan Road, Beijing 100049, China}

\date{\today}

\begin{abstract}

We try to reconcile the tension between the local 2.4\% determination of Hubble constant and its global determination by $\textit{Planck}$ CMB data and BAO data through modeling the dark energy variously. We find that the chi-square is significantly reduced by $\Delta\chi^2_\text{all}=-6.76$ in the redshift-binned dark energy model where the $68\%$ limits of the equation of state of dark energy read $w(0\leq z\leq 0.1)=-1.958_{-0.508}^{+0.509}$, $w(0.1< z\leq 1.5)=-1.006_{-0.082}^{+0.092}$, and here $w(z>1.5)$ is fixed to $-1$.

\end{abstract}


\maketitle


\section{Introduction}

Recently Riess et al. in \cite{Riess:2016jrr} confirmed and improved their former determination of the Hubble constant ($H_0=73.8\pm2.4~\textrm{km}~\textrm{s}^{-1}~\textrm{Mpc}^{-1}$ \cite{Riess:2011yx}): 
\m
H_0=73.00\pm1.75~\textrm{km}~\textrm{s}^{-1}~\textrm{Mpc}^{-1}
\n 
at $68\%$ confidence level (C.L.) by still using three distance anchors (Milky Way (MW), Large Magellic Cloud (LMC) and NGC 4258). The improvement in uncertainty is due to not only an enlarged number of SN hosts, Cepheids (in the LMC) and MW Cepheids (with \textit{HST}-based trigonometric parallaxes) but also a revised distance to NGC 4258 and the LMC. In 2013 Efstathiou  revisited the dataset of \cite{Riess:2011yx} and yielded another two different value of $H_0$: $H_0=72.5\pm2.5~\textrm{km}~\textrm{s}^{-1}~\textrm{Mpc}^{-1}$ using three anchors (MW, LMC and NGC 4258) and $H_0=70.6\pm3.3~\textrm{km}~\textrm{s}^{-1}~\textrm{Mpc}^{-1}$ with NGC 4258 as only anchor in \cite{Efstathiou:2013via}. As emphasized in \cite{Riess:2016jrr}, although these three local determinations of $H_0$ with the same three anchors are consistent with each other, $H_0=73.00\pm1.75~\textrm{km}~\textrm{s}^{-1}~\textrm{Mpc}^{-1}$ is considered to be the best one (R16, hereafter) due to the aforementioned improvement.

On the other hand, there are some widely accepted global determinations of $H_0$ mainly derived from CMB data under the assumption of $\Lambda$CDM model:
$H_0=69.7\pm2.1~\textrm{km}~\textrm{s}^{-1}~\textrm{Mpc}^{-1}$ from WMAP9 \cite{Ade:2015xua}
\footnote{The constraint from WMAP9 is different the original ones in \cite{Bennett:2012zja,Hinshaw:2012aka} which does not include $0.06~\textrm{eV}$ neutrino mass. };
$H_0=68.0\pm0.7~\textrm{km}~\textrm{s}^{-1}~\textrm{Mpc}^{-1}$ from WMAP9+BAO \cite{Ade:2015xua};
$H_0=69.3\pm0.7~\textrm{km}~\textrm{s}^{-1}~\textrm{Mpc}^{-1}$ from WMAP9+ACT+SPT+BAO \cite{Bennett:2014tka};
$H_0=67.3\pm1.0~\textrm{km}~\textrm{s}^{-1}~\textrm{Mpc}^{-1}$ from \textit{Planck} TT+lowP \cite{Ade:2015xua};
$H_0=67.6\pm0.6~\textrm{km}~\textrm{s}^{-1}~\textrm{Mpc}^{-1}$ from  \textit{Planck} TT+lowP+BAO \cite{Ade:2015xua}. 
In addition, in \cite{Cheng:2014kja} combining the low-redshift and high-redshift isotropic BAO data, the Hubble constant was given by $H_0=68.17_{-1.56}^{+1.55}~\textrm{km}~\textrm{s}^{-1}~\textrm{Mpc}^{-1}$ which is consistent with the results from CMB data. 
Comparing R16 with these cosmological estimates, we find that there is a longstanding tension which becomes more significant now.

Since the global determinations of $H_0$ are highly model-dependent, there is a well known solution to this tension: adding additional dark radiation to the base $\Lambda$CDM model \cite{Ade:2013zuv,Ade:2015xua,Riess:2016jrr,Mehta:2012hh,Cheng:2013csa}. Recently in \cite{DiValentino:2016hlg} the authors found that a variation of equation of state (EOS) of Dark Energy (DE) in a 12-parameter extension of $\Lambda$CDM model is more favored than adding dark radiation. Furthermore, they pointed out that the tension between R16 and the combination of \textit{Plank} 2015 data and BAO data still exists in their 12-parameter extension. In this short paper we try to reconcile the tension between R16 with the combination of $\textit{Planck}$ CMB data and BAO data through modeling the DE variously.

In the literatures there are several well-known DE models: the $\Lambda$CDM model where DE is described by a cosmological constant with EOS fixed as $-1$; the $w$CDM model where DE has a constant EOS $w$; the CPL model in which EOS of DE is time-dependent \cite{Chevallier:2000qy,Linder:2002et}. In general the non-parametric reconstructions of the EOS of DE in bins of redshift, like those in \cite{Huang:2009rf,Li:2011wb,Huang:2015vpa}, are considered to be model-independent. We will investigate whether these DE models can be used to reconcile the tension on the determinations of Hubble constant in this short paper. 

The rest of the paper is arranged as follows. In Sec.~\ref{dataset}, we reveal our methodology and cosmological data sets used in this paper. In Sec.~\ref{results}, we globally fit all of the cosmological parameters in various extensions to the $\Lambda$CDM model by combining R16, $\textit{Planck}$ and BAO datasets. A brief summary and discussion are included in Sec.~\ref{conclusion}.

\section{Data and Methodology}\label{dataset}

In this paper, we try to reconcile the tension between R16 and the global determination of $H_0$ by CMB data released by $\textit{Planck}$ collaboration in 2015 and BAO measurements. Here we add the R16 prior to the data combination of $\textit{Planck}$~2015 data ($\textit{Planck}$ TT,TE,EE+lowP+lensing) \cite{Ade:2015xua} and the BAO data including 6dFGS  ($z_{\textrm{eff}}=0.106$) \cite{Beutler:2011hx}, MGS ($z_{\textrm{eff}}=0.15$) \cite{Ross:2014qpa}, BOSS DR12 LOWZ ($z_{\textrm{eff}}=0.32$) \cite{Cuesta:2015mqa} and CMASS ($z_{\textrm{eff}}=0.57$) \cite{Cuesta:2015mqa}.

We focus on a spatially flat universe and the Friedmann equation reads \begin{equation}
\label{H}
H^2=\frac{8\pi G}{3}\left[\rho_{r}(0)(1+z)^4+\rho_{m}(0)(1+z)^3+\rho_{de}(z)\right],
\end{equation}
where the energy density of DE is related to its EOS $(w(z))$ by  
\begin{equation}
\label{rho}
\rho_{de}(z)=\rho_{de}(0)~\textrm{exp}\left\{3\int_0^z\frac{dz'}{(1+z')}\left[1+w(z')\right]\right\}. 
\end{equation}
The global determination of $H_0$ highly depends on the evolution of the energy density at low redshift, and consequently on the evolution of EOS of DE at low redshift.

In the $\Lambda$CDM model, $w(z)=-1$ and there are six base cosmological parameters which are denoted by \{$\Omega_bh^2$,$\Omega_ch^2$,$100\theta_{\textrm{MC}}$,$\tau,n_s$,$\textrm{ln}(10^{10}A_s)$\}. Here $\Omega_bh^2$ is the physical density of baryons today, $\Omega_ch^2$ is the physical density of cold dark matter today, $\theta_{\textrm{MC}}$ is the ratio between the sound horizon and the angular diameter distance at the decoupling epoch, $\tau$ is the Thomson scatter optical depth due to reionization, $n_s$ is the scalar spectrum index and $A_s$ is the amplitude of the power spectrum of primordial curvature perturbations at the pivot scale $k_p=0.05$ Mpc$^{-1}$.

The $w$CDM model with an arbitrary constant EOS $w$ is the simplest DE extension of $\Lambda$CDM model. Therefore, there are six base cosmological parameters plus another free parameter $w$. 
The CPL model is a widely used DE extension of $\Lambda$CDM model and the EOS of DE is parametrized by 
\begin{equation}
\label{CPL}
w(z)=w_0+w_a\frac{z}{1+z},
\end{equation}
where $w_0$ and $w_a$ are two free parameters in addition to the six base free parameters mentioned above. Usually we use $w_0w_a$CDM to donate this model.

Furthermore, we also consider two model-independent redshift-binned DE models. 
Since the expansion of the universe is insensitive to DE at high redshifts where the energy density of DE become negligibly small, we fix the EOS of DE to be $-1$ for $z>1.5$ for convenience. The first redshift-binned DE model is denoted by $w_{0.25}w_{1.5}$CDM in which the EOS is divided into three bins as follows 
\begin{align}
w(z)= 
\begin{cases}
w_{0.25}, ~~z\leq 0.25\ ;\\
w_{1.5}, ~~0.25<z\leq 1.5\ ;\\
-1, ~~z>1.5\ ,
\end{cases}
\end{align}
where $w_{0.25}$ and $w_{1.5}$ are two free parameters. 
On the other hand, we notice that the effective redshift for the lowest-redshift BAO is $z_{\textrm{eff}}=0.106$. In order to ``effectively" exclude the distance scale constraints from BAO datasets, we also consider another redshift-binned DE model ($w_{0.1}w_{1.5}$CDM model) in which the EOS of DE is given by 
\begin{align}
w(z)=
\begin{cases}
w_{0.1}, ~~z\leq 0.1\ ;\\
w_{1.5}, ~~0.1<z\leq 1.5\ ;\\
-1, ~~z>1.5\ ,
\end{cases}
\end{align}
where $w_{0.1}$ and $w_{1.5}$ are two free parameters. In this case, there is no BAO dataset in the first redshift bin ($0\leq z\leq 0.1$).

\section{Results}\label{results}

Our main results are given in Tab.~\ref{tab:results} where we represent the $68\%$ limits for the cosmological parameters in different DE extensions to $\Lambda$CDM model from the data combination of $\textit{Planck}$ TT,TE,EE+lowP+lensing+BAO+R16. 
\begin{table*}[!htp]
\centering
\renewcommand{\arraystretch}{1.5}
\scalebox{1.0}[1.0]{%
\begin{tabular}{c|c c c c  c}
\hline
--&$\Lambda$CDM    &$w$CDM    &$w_0w_a$CDM    &$w_{0.1}w_{1.5}$CDM    &$w_{0.25}w_{1.5}$CDM\\
\hline
$\Omega_bh^2$
&$0.02236_{-0.00015}^{+0.00014}$
&$0.02223_{-0.00016}^{+0.00015}$
&$0.02222\pm0.00015$
&$0.02227\pm0.00015$
&$0.02225_{-0.00016}^{+0.00015}$    \\
$\Omega_ch^2$
&$0.1180_{-0.0010}^{+0.0011}$
&$0.1197_{-0.0013}^{+0.0012}$
&$0.1196\pm0.0013$
&$0.1191_{-0.0014}^{+0.0013}$
&$0.1193\pm0.0013$      \\
$100\theta_{\emph{MC}}$
&$1.04101\pm0.00029$
&$1.04082_{-0.00030}^{+0.00031}$
&$1.04082_{-0.00032}^{+0.00031}$
&$1.04089\pm0.00031$
&$1.04086_{-0.00031}^{+0.00030}$    \\
$\tau$
&$0.071\pm0.012$
&$0.057_{-0.014}^{+0.013}$
&$0.058\pm0.015$
&$0.064\pm0.014$
&$0.061\pm0.014$        \\
${\textrm{ln}}(10^{10}A_s)$
&$3.072\pm0.023$
&$3.047\pm0.025$
&$3.050\pm0.028$
&$3.060_{-0.026}^{+0.027}$
&$3.055_{-0.025}^{+0.026}$        \\
$n_s$
&$0.9686\pm0.0041$
&$0.9643\pm0.0044$
&$0.9643_{-0.0044}^{+0.0045}$
&$0.9659\pm0.0046$
&$0.9652\pm0.0045$      \\
EOS$(z=0)$
&--
&$w=-1.113_{-0.055}^{+0.056}$
&$w_0=-1.185_{-0.211}^{+0.185}$
&$w_{0.1}=-1.958_{-0.508}^{+0.509}$
&$w_{0.25}=-1.296_{-0.202}^{+0.203}$\\
--
&--
&--
&$w_a=0.196_{-0.485}^{+0.664}$
&$w_{1.5}=-1.006_{-0.082}^{+0.092}$
&$w_{1.5}=-1.037_{-0.113}^{+0.112}$\\
${H_0}({\textrm{km}\cdot\textrm{s}^{-1}\cdot\textrm{Mpc}^{-1}})$
&$68.08_{-0.48}^{+0.47}$
&$70.71_{-1.49}^{+1.35}$
&$71.29_{-1.88}^{+1.90}$
&$74.18_{-2.51}^{+2.54}$
&$72.07_{-2.01}^{+1.98}$         \\
$\Omega_{\Lambda}$
&$0.6957\pm0.0062$
&$0.7146\pm0.0107$
&$0.7190_{-0.0145}^{+0.0163}$
&$0.7411_{-0.0169}^{+0.0199}$
&$0.7256_{-0.0146}^{+0.0171}$\\
$\Omega_m$
&$0.3043\pm0.0062$
&$0.2854\pm0.0107$
&$0.2810_{-0.0163}^{+0.0145}$
&$0.2589_{-0.0199}^{+0.0169}$
&$0.2744_{-0.0171}^{+0.0146}$\\
\hline
$\chi^2_{\textrm{all}}$
&12966.04     &12962.42     &12962.26    &12959.28     &12962.28\\
$\chi^2_{\textrm{CMB}}$
&12957.07     &12955.64     &12955.50    &12954.74     &12956.44\\
$\chi^2_{\textrm{BAO}}$
&2.40         &4.38         &5.10        &4.48         &5.24    \\
$\chi^2_{H_0}$
&6.57         &2.40         &1.66        &0.06         &0.60    \\
\hline
\end{tabular}}
\caption{The $68\%$ limits for the cosmological parameters in different DE extensions to $\Lambda$CDM model from the data combination of $\textit{Planck}$ TT,TE,EE+lowP+lensing+BAO+R16. The $\chi^2$ for different models against individual data are also listed explicitly.}
\label{tab:results}
\end{table*}

Combining R16 with the $\textit{Planck}$~2015 and BAO data, the constraints on the six parameters in the $\Lambda$CDM model almost do not change with respect to the case without adding a local $H_0$ prior, but $\chi^2_{H_0}=6.57$ indicates the existence of tension between the local and global determinations of $H_0$. For the $\Lambda$CDM model, $\chi^2_{\rm BAO}=2.40$ implies that the $\textit{Planck}$~2015 and BAO data are consistent with each other.

Compared to the $\Lambda$CDM model, the chi-square is reduced by $\Delta\chi^2_{\textrm{all}}=-3.62$ in $w$CDM model, and a phantom-like DE with EOS of $w=-1.113_{-0.055}^{+0.056}$ is preferred at more than $95\%$ C.L.. However, in $w$CDM model, the Hubble constant reads $70.71_{-1.49}^{+1.35}$ $\text{km}\cdot \text{s}^{-1}\cdot \text{Mpc}^{-1}$ which is still small compared to the local determination.

\begin{figure}[]
\begin{center}
\includegraphics[width=\figurewidth]{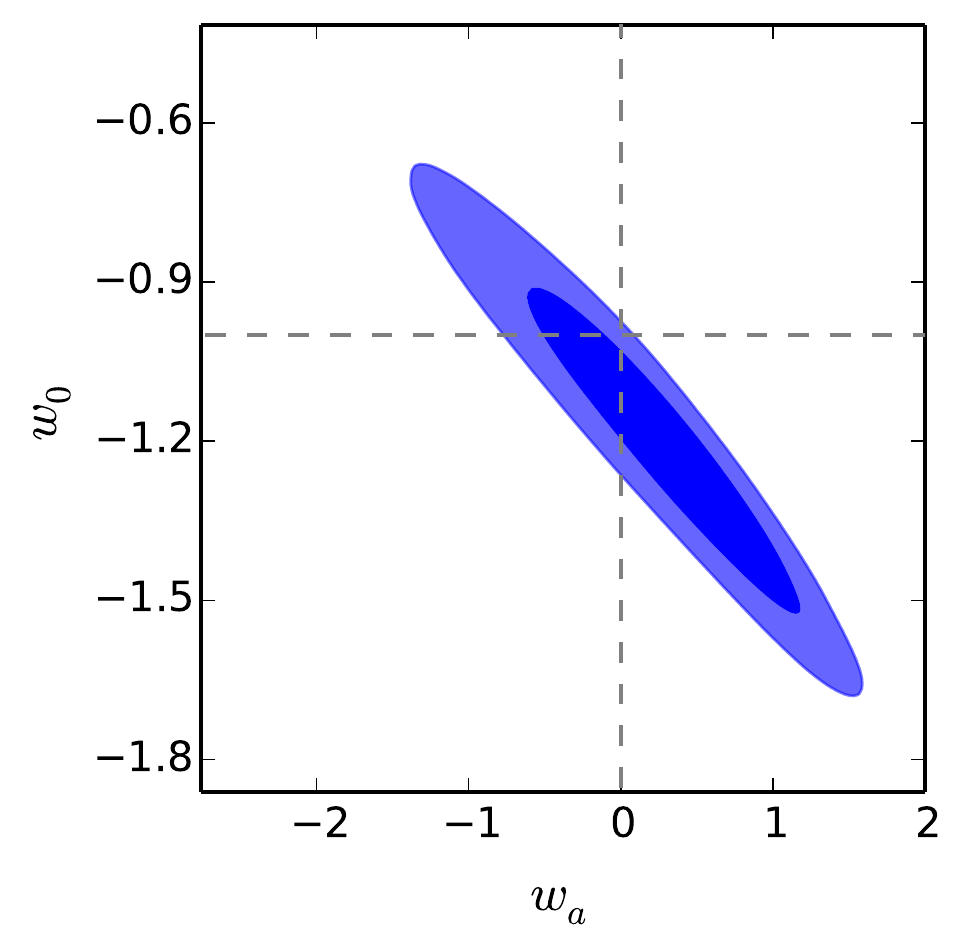}
\end{center}
\caption{The contour plot of $w_0$ and $w_a$ in $w_0w_a$CDM model.}
\label{fig:w0waCDM}
\end{figure}

Even though a parameter $w_a$ for describing the evolution of the EOS of DE is included in the $w_{0}w_{a}$CDM model, the chi-square is almost the same as that in $w$CDM model. The contour plot of $w_0$ and $w_a$ shows up in Fig.~\ref{fig:w0waCDM}. 
From Fig.~\ref{fig:w0waCDM}, the cosmological constant is disfavored at around $95\%$ C.L., and $w_0=-1.185_{-0.211}^{+0.185}$ is consistent with the constraint on $w$ in $w$CDM model. 
Actually, the $w_{0}w_{a}$CDM model provides a little bit better fit to the Hubble constant ($\chi^2_{H_0}=1.66$, and $\Delta \chi^2_{H_0}=-0.74$ compared to $w$CDM model), but the fitting to the BAO datasets becomes worse by $\Delta\chi^2_\text{BAO}=0.72$. Thus, from the statistic point of view, the data do not prefer this model.

\begin{figure}[]
\begin{center}
\includegraphics[width=\figurewidth]{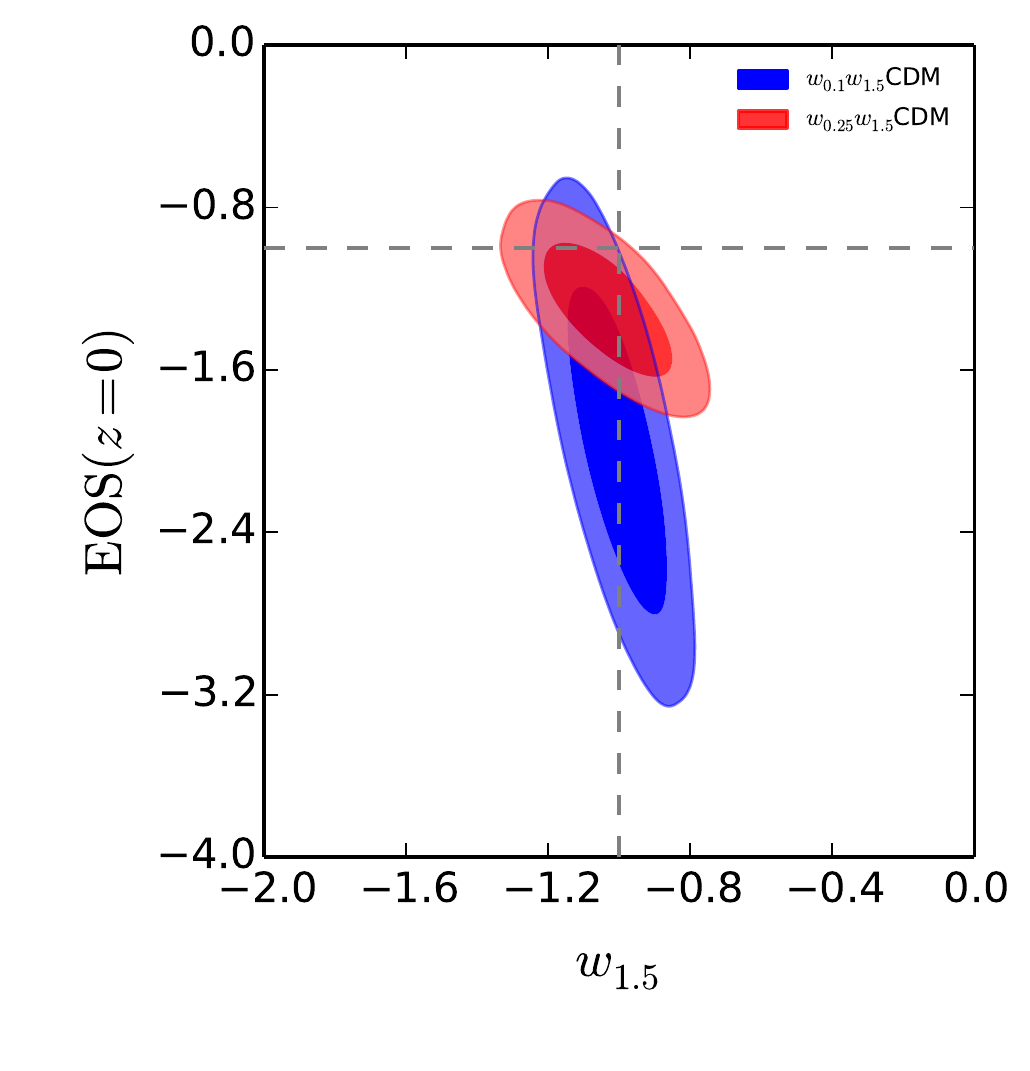}
\end{center}
\caption{The plots of EOS at $z=0$ and $w_{1.5}$ in $w_{0.1}w_{1.5}$CDM model and $w_{0.25}w_{1.5}$CDM model.}
\label{fig:2wwCDM}
\end{figure}

It is interesting to see that the chi-square for $w_{0.1}w_{1.5}$CDM model is significantly reduced by $\Delta\chi^2_\text{all}=-6.76$ compared to $\Lambda$CDM model, and $\Delta\chi^2_\text{all}=-3.14$ compared to $w$CDM model. 
Furthermore, $H_0=74.18_{-2.51}^{+2.54}$ $\text{km}\cdot \text{s}^{-1}\cdot \text{Mpc}^{-1}$ and $\chi^2_{H_0}=0.06$ indicate that the global fitting is consistent with the local determination of Hubble constant in $w_{0.1}w_{1.5}$CDM model. However, the fitting becomes worse in $w_{0.25}w_{1.5}$CDM model ($\Delta \chi^2_\text{all}=3.00$, $\Delta \chi^2_\text{CMB}=1.70$, $\Delta \chi^2_\text{BAO}=0.76$ and $\Delta \chi^2_{H_0}=0.54$ compared to $w_{0.1}w_{1.5}$CDM model).
That is to say, all of the individual datasets prefer $w_{0.1}w_{1.5}$CDM model compared to $w_{0.25}w_{1.5}$CDM model. 
The plots of EOS at $z=0$ and $w_{1.5}$ in $w_{0.1}w_{1.5}$CDM model and $w_{0.25}w_{1.5}$CDM model are illustrated in Fig.~\ref{fig:2wwCDM}. 
We see that the cosmological constant is disfavored at more than $95\%$ C.L. in $w_{0.1}w_{1.5}$CDM model.
Notice that there are two BAO data points in the range of $0.1<z<0.25$, and they are consistent with $\textit{Planck}$ CMB data in the $\Lambda$CDM model. But the combination of the local determination of Hubble constant and the $\textit{Planck}$ CMB data prefers a phantom-like DE model. Therefore the EOS of DE $w_{0.25}<-1$ in the range of $0<z<0.25$ enhances the chi-squares in $w_{0.25}w_{1.5}$CDM model compared to $w_{0.1}w_{1.5}$CDM model.

\section{Summary and discussion}\label{conclusion}

To summarize, the tension between the local determination of Hubble constant and the global determination by  $\textit{Planck}$ CMB data and BAO data is statistically significant in the $\Lambda$CDM model, and the EOS of DE is preferred to be less than -1 at low redshifts when R16 is added to $\textit{Planck}$ CMB and BAO data. The chi-square for the $w_{0.1}w_{0.25}$CDM model is significantly reduced by $\Delta\chi^2_\text{all}=-6.76$ compared to the $\Lambda$CDM model, and this model can reconcile the tension on determination of Hubble constant between R16 and the combination of $\textit{Planck}$ CMB data and BAO data.

\begin{figure}[]
\begin{center}
\includegraphics[width=\figurewidth]{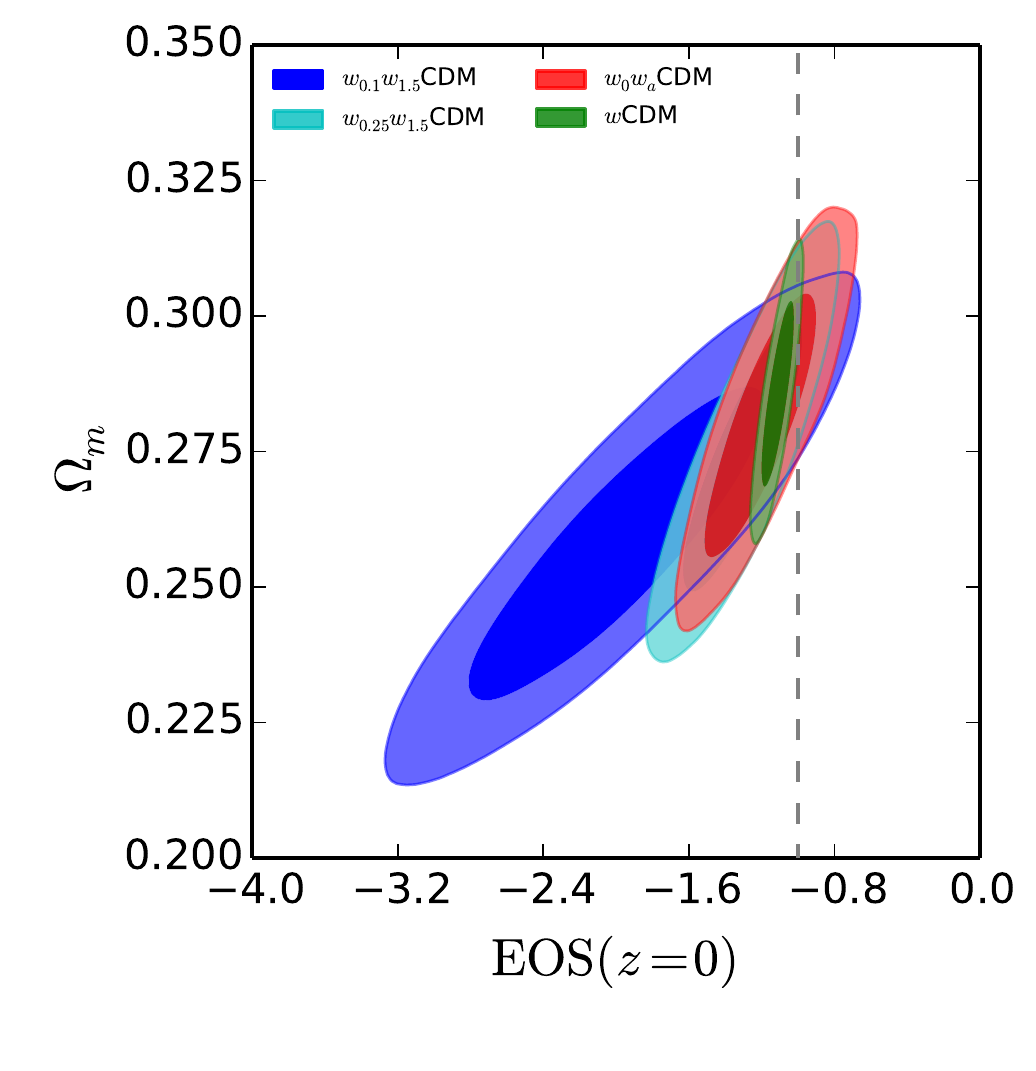}
\end{center}
\caption{The contour plot of $\Omega_m$ and EOS at $z=0$ for the models extended to the $\Lambda$CDM model.}
\label{fig:extCDM}
\end{figure}

\begin{figure}[]
\begin{center}
\includegraphics[width=\figurewidth]{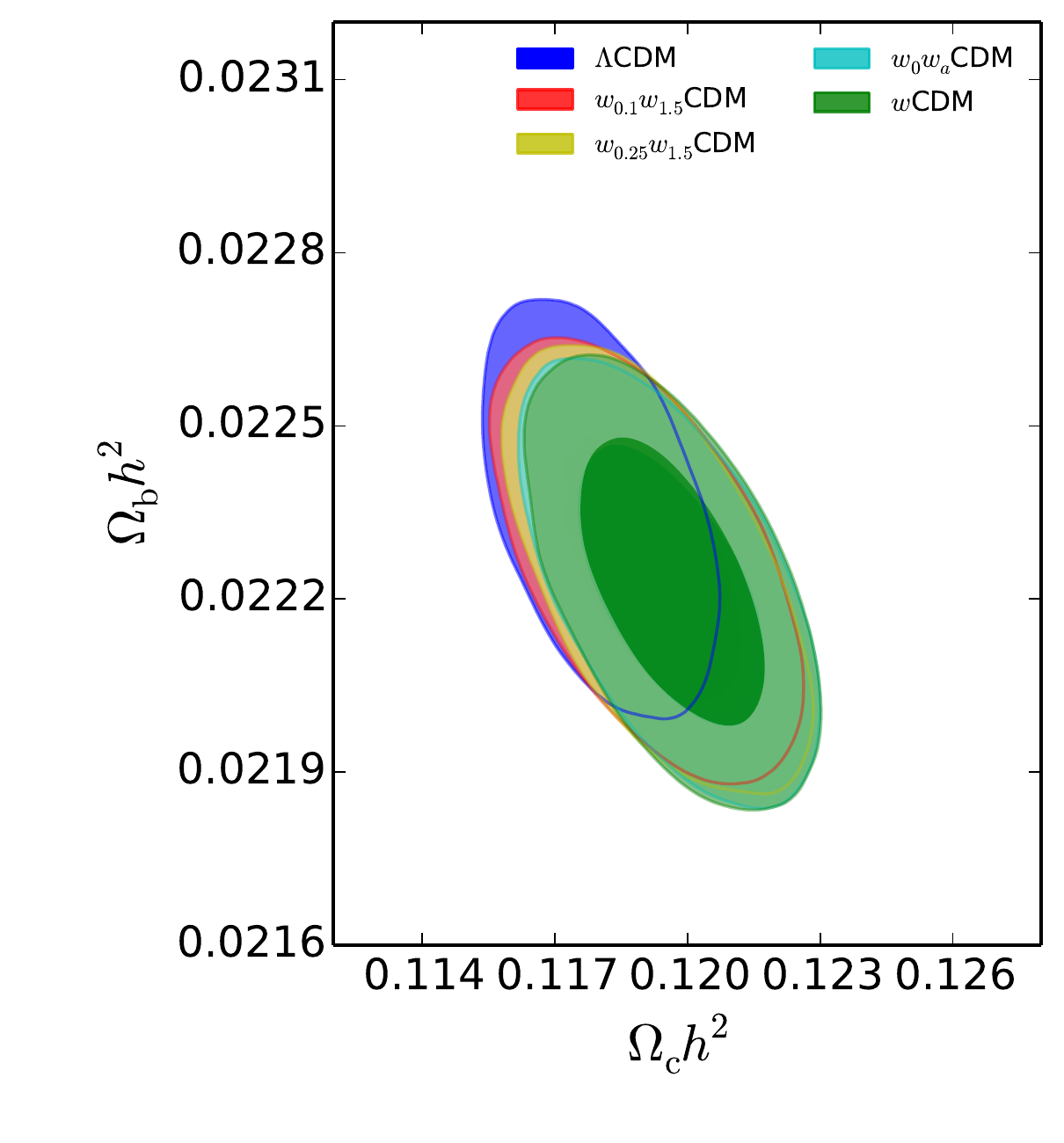}
\end{center}
\caption{Constraints on $\Omega_bh^2$ and $\Omega_ch^2$ from different DE extension of $\Lambda$CDM model.}
\label{fig:bc}
\end{figure}

Since the EOS of DE is preferred to be less than -1 at low redshifts, the matter energy density today becomes smaller than that in the $\Lambda$CDM model. See Fig.~\ref{fig:extCDM}. 
In particular, $\Omega_m=0.2589_{-0.0199}^{+0.0169}$ in the $w_{0.1}w_{0.25}$CDM model, and it is much lower than that in the $\Lambda$CDM model ($\Omega_m=0.3043\pm 0.0062$). 
One may worry that this model cannot fit the $\textit{Planck}$ CMB data because of the lower matter density today. Actually CMB power spectra are more sensitive to $\Omega_bh^2$ and $\Omega_ch^2$, not $\Omega_m$. The constraints on $\Omega_bh^2$ and $\Omega_ch^2$ for different models are showed in Fig.~\ref{fig:bc} from which we see that the constraints on $\Omega_bh^2$ and $\Omega_ch^2$ in all of these models are similar to each other.

Finally, with reduction of large-scale systematic effects in HFI polarization maps, \textit{Planck} collaboration gave some new constraints on the reinoization optical depth in \cite{Aghanim:2016yuo,Adam:2016hgk} recently. For example, for the likelihood method of Lollipop and the conservative cross-spectra estimators of PCL, the constraint on the reinoization optical depth was $\tau=0.053_{-0.016}^{+0.011}$ which is smaller than that given by \textit{Planck} TT and lowP power spectra. However, the reionization optical depth in the models extended to the $\Lambda$CDM models in this paper prefer a lower reionization optical depth which is more consistent with the recent result released by the \textit{Planck} collaboration. 

In a word, if all of the datasets including the local determination of Hubble constant, \textit{Planck} CMB data and BAO data are reliable, it should imply some new physics beyond the six-parameter base $\Lambda$CDM model.

\noindent {\bf Acknowledgments}

We would like to thank Sai Wang for the helpful conversations. 
This work is supported by Top-Notch Young Talents Program of China and grants from NSFC (grant NO. 11322545, 11335012 and 11575271).




\begin{thebibliography}{99}
\frenchspacing

\bibitem{Riess:2016jrr}
  A.~G.~Riess {\it et al.},
  arXiv:1604.01424 [astro-ph.CO].  

\bibitem{Riess:2011yx}
  A.~G.~Riess {\it et al.},
  Astrophys.\ J.\  {\bf 730}, 119 (2011)  Erratum: [Astrophys.\ J.\  {\bf 732}, 129 (2011)]  doi:10.1088/0004-637X/732/2/129, 10.1088/0004-637X/730/2/119  [arXiv:1103.2976 [astro-ph.CO]].  

\bibitem{Efstathiou:2013via}
  G.~Efstathiou,
  Mon.\ Not.\ Roy.\ Astron.\ Soc.\  {\bf 440}, no. 2, 1138 (2014)
  doi:10.1093/mnras/stu278  [arXiv:1311.3461 [astro-ph.CO]].  

\bibitem{Ade:2015xua}
  P.~A.~R.~Ade {\it et al.} [Planck Collaboration],
  arXiv:1502.01589 [astro-ph.CO].  


\bibitem{Bennett:2012zja}
  C.~L.~Bennett {\it et al.} [WMAP Collaboration],
  Astrophys.\ J.\ Suppl.\  {\bf 208}, 20 (2013)  doi:10.1088/0067-0049/208/2/20  [arXiv:1212.5225 [astro-ph.CO]].  

\bibitem{Hinshaw:2012aka}
  G.~Hinshaw {\it et al.} [WMAP Collaboration],
  Astrophys.\ J.\ Suppl.\  {\bf 208}, 19 (2013)  doi:10.1088/0067-0049/208/2/19  [arXiv:1212.5226 [astro-ph.CO]].  

\bibitem{Bennett:2014tka}
  C.~L.~Bennett, D.~Larson, J.~L.~Weiland and G.~Hinshaw,
  Astrophys.\ J.\  {\bf 794}, 135 (2014)  doi:10.1088/0004-637X/794/2/135  [arXiv:1406.1718 [astro-ph.CO]].  

\bibitem{Cheng:2014kja} 
  C.~Cheng and Q.~G.~Huang,
  Sci.\ China Phys.\ Mech.\ Astron.\  {\bf 58}, no. 9, 599801 (2015)
  doi:10.1007/s11433-015-5684-5
  [arXiv:1409.6119 [astro-ph.CO]].

\bibitem{Ade:2013zuv}
  P.~A.~R.~Ade {\it et al.} [Planck Collaboration],
  Astron.\ Astrophys.\  {\bf 571}, A16 (2014)  doi:10.1051/0004-6361/201321591  [arXiv:1303.5076 [astro-ph.CO]].  


\bibitem{Mehta:2012hh}
  K.~T.~Mehta, A.~J.~Cuesta, X.~Xu, D.~J.~Eisenstein and N.~Padmanabhan,
  Mon.\ Not.\ Roy.\ Astron.\ Soc.\  {\bf 427}, 2168 (2012)  doi:10.1111/j.1365-2966.2012.21112.x  [arXiv:1202.0092 [astro-ph.CO]].  


\bibitem{Cheng:2013csa} 
  C.~Cheng and Q.~G.~Huang,
  Phys.\ Rev.\ D {\bf 89}, no. 4, 043003 (2014)
  doi:10.1103/PhysRevD.89.043003
  [arXiv:1306.4091 [astro-ph.CO]].


\bibitem{DiValentino:2016hlg}
  E.~Di Valentino, A.~Melchiorri and J.~Silk,
  arXiv:1606.00634 [astro-ph.CO].  



\bibitem{Chevallier:2000qy} 
  M.~Chevallier and D.~Polarski,
  Int.\ J.\ Mod.\ Phys.\ D {\bf 10}, 213 (2001)
  doi:10.1142/S0218271801000822
  [gr-qc/0009008].
  
\bibitem{Linder:2002et} 
  E.~V.~Linder,
  Phys.\ Rev.\ Lett.\  {\bf 90}, 091301 (2003)
  doi:10.1103/PhysRevLett.90.091301
  [astro-ph/0208512].  


\bibitem{Huang:2009rf}
  Q.~G.~Huang, M.~Li, X.~D.~Li and S.~Wang,
  Phys.\ Rev.\ D {\bf 80}, 083515 (2009)
  [arXiv:0905.0797 [astro-ph.CO]].

\bibitem{Li:2011wb}
  X.~D.~Li, S.~Li, S.~Wang, W.~S.~Zhang, Q.~G.~Huang and M.~Li,
  JCAP {\bf 1107}, 011 (2011)
  [arXiv:1106.4116 [astro-ph.CO]].

\bibitem{Huang:2015vpa}
  Q.~G.~Huang, K.~Wang and S.~Wang,
  JCAP {\bf 1512}, no. 12, 022 (2015)  doi:10.1088/1475-7516/2015/12/022  [arXiv:1509.00969 [astro-ph.CO]].  

\bibitem{Beutler:2011hx}
  F.~Beutler {\it et al.},
  Mon.\ Not.\ Roy.\ Astron.\ Soc.\  {\bf 416}, 3017 (2011)
  doi:10.1111/j.1365-2966.2011.19250.x
  [arXiv:1106.3366 [astro-ph.CO]].

\bibitem{Ross:2014qpa}
  A.~J.~Ross, L.~Samushia, C.~Howlett, W.~J.~Percival, A.~Burden and M.~Manera,
  Mon.\ Not.\ Roy.\ Astron.\ Soc.\  {\bf 449}, no. 1, 835 (2015)
  doi:10.1093/mnras/stv154
  [arXiv:1409.3242 [astro-ph.CO]].
  doi:10.1093/mnras/stu523
  [arXiv:1312.4877 [astro-ph.CO]].

\bibitem{Cuesta:2015mqa}
  A.~J.~Cuesta {\it et al.},
  arXiv:1509.06371 [astro-ph.CO].




\bibitem{Aghanim:2016yuo}
  N.~Aghanim {\it et al.} [Planck Collaboration],
  arXiv:1605.02985 [astro-ph.CO].  

\bibitem{Adam:2016hgk}
  R.~Adam {\it et al.} [Planck Collaboration],
  arXiv:1605.03507 [astro-ph.CO].  










\end{thebibliography}
\end{document}